\documentclass{PoS}
\usepackage{graphicx}

\title{OH masers in the Milky Way and Local Group galaxies in the SKA era}

\ShortTitle{OH masers}

\author{\speaker{Sandra Etoka}$^1$, Dieter Engels$^1$, Hiroshi Imai$^2$, 
Joanne Dawson$^3$, Simon Ellingsen$^4$, Lorant Sjouwerman$^5$, 
Huib van~Langevelde$^{6,7}$ \\
	$^1$Hamburger Sternwarte, $^2$Kagoshima University, 
        $^3$Macquarie University, $^4$University of Tasmania, 
        $^5$National Radio Astronomy Observatory,
        $^6$Joint Institute for VLBI in Europe, $^7$Sterrewacht Leiden\\
        E-mail: \email{sandra.etoka@googlemail.com}
        }

\abstract{The intense line emission of OH masers is a perfect tracer of regions
where new stars are born as well as of evolved stars, shedding large
amounts of processed matter into the interstellar medium. From SKA
deep surveys at 18~cm, where the maser lines from the ground-state of
the OH molecule arise, we predict the discovery of more than 
20000 sources of stellar and interstellar origin throughout the Galaxy. 
The study of this maser emission has many applications, including the 
determination of magnetic field strengths from polarisation measurements, 
studies of stellar kinematics using the precisely determined radial 
velocities, and distance determinations from VLBI astrometry. 
A new opportunity to study shocked gas in different galactic environments 
is expected to arise with the detection of lower luminosity masers. For the 
first time, larger numbers of OH masers will be detected in Local Group
galaxies. New insights are expected in structure formation in galaxies
by comparing maser populations in galaxies of different metallicity,
as both their properties as well as their numbers depend on it. With
the full capabilities of SKA, further maser transitions such as from
excited OH and from methanol will be accessible, providing new tools
to study the evolution of star-forming regions in particular.
}

\FullConference{
Advancing Astrophysics with the Square Kilometre Array\\
June 8-13, 2014\\
Giardini Naxos, Italy}

\newcommand{\skipthis}[1]{}

\newcommand\water{H$_2$O}
\newcommand\methanol{CH$_3$OH}

\newcommand{\kms}{km\,s$^{-1}$}

\newcommand{\Myr}{$M_{\odot}$\,yr$^{-1}$}
\newcommand{\Lsun}{$L_{\odot}$}
\begin{document}

\section{Introduction}
\label{Introduction}
In the Milky Way, maser emission is often observed in the
circumstellar shells of red giant stars and in the surroundings of
young stellar objects (YSOs). These are environments which are cool enough to
form molecules and provide sufficient velocity coherence and density,
so that the masers are naturally excited. The strongest stellar masers
are those of oxygen-bearing molecules like hydroxyl (OH), water
(\water) and silicon-monoxide (SiO). In addition, methanol (\methanol) 
and to lesser extent, ammonia (NH$_3$) and formaldehyde (H$_2$CO) masers 
are detected close to star-formation regions (SFRs). The major maser 
transitions accessible with the SKA during Phase 1 will be from OH at 
frequencies of 1612, 1665 and 1667~MHz in band~2 and 1720~MHz in band~3.

\begin{figure}[h] 
\begin{minipage}[t]{15cm}
  \begin{minipage}[t]{7.5cm}
   \begin{flushleft}
      \hspace*{-0.4cm}{\includegraphics*[width=6cm,angle=-90]{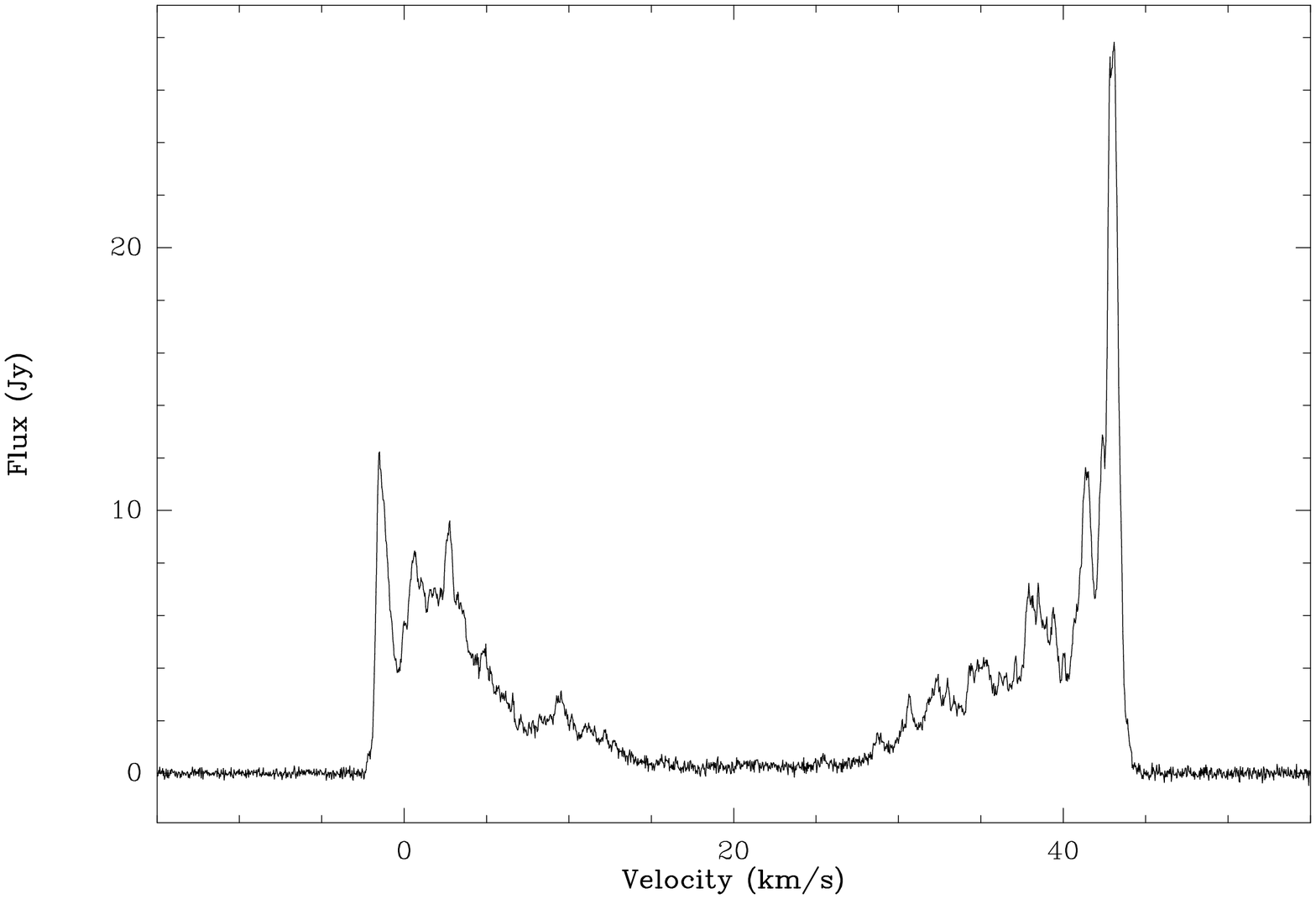}}
   \end{flushleft}
  \end{minipage}
  \begin{minipage}[t]{7.5cm}
   \begin{flushright}
      \hspace*{-0.4cm}{\includegraphics*[width=6cm,angle=-90]{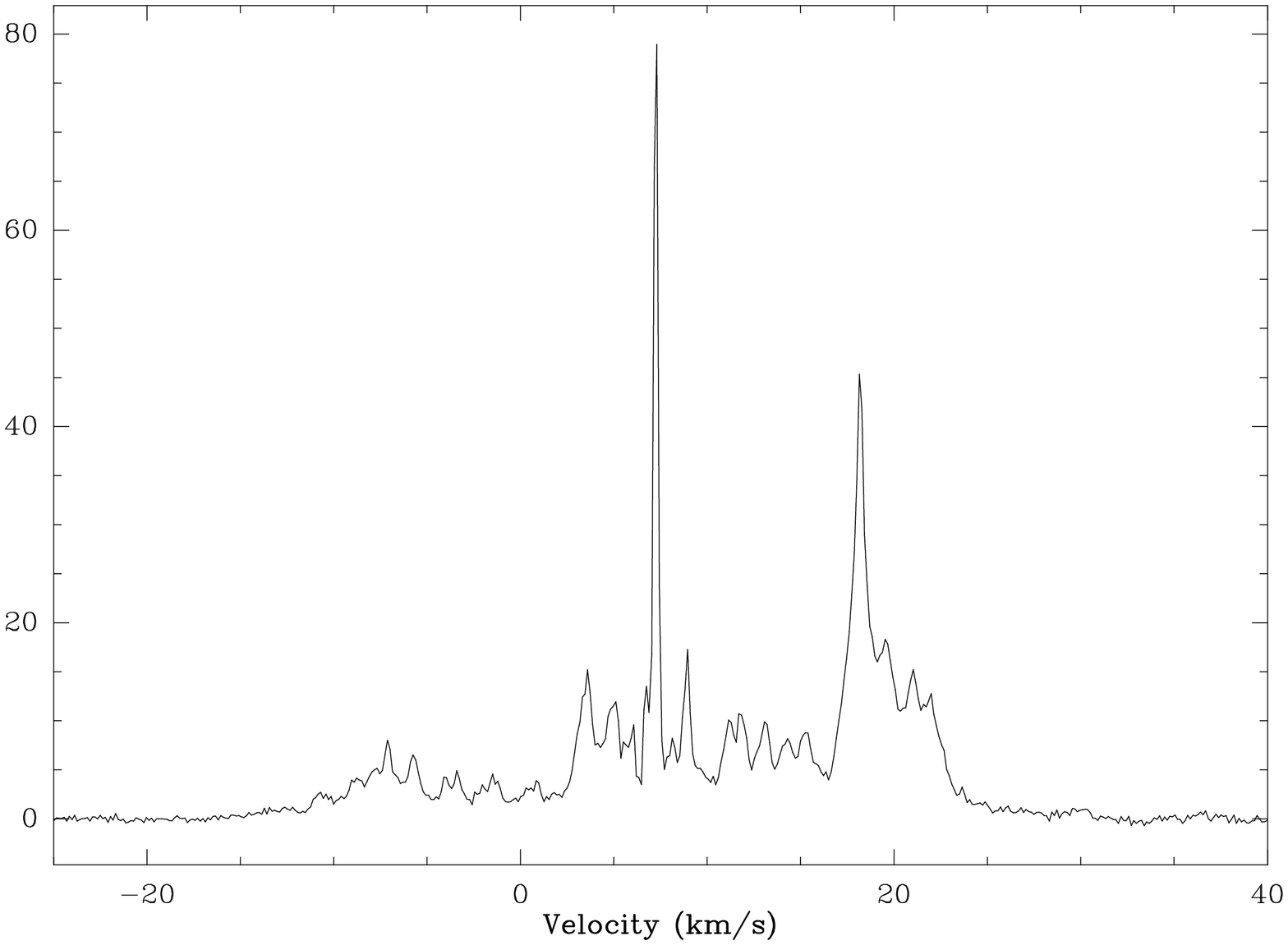}}
   \end{flushright}
  \end{minipage}
\end{minipage}
\caption{\small {\bf Left:} 1612-MHz OH maser of the long-period 
         variable star OH\,16.1-0.2 {\bf Right: } 1665-MHz OH maser of the 
         star-forming region Orion~KL. Both spectra were taken with the 
         {Nan\c cay} radiotelescope.}
\label{oh_spectra}
\end{figure}

Known OH masers in the Galaxy are typically located in the plane at
distances 2-8 kpc. The evolved-star and star-forming-region origin can be 
differentiated from their spectral shapes and the ratio of intensities 
between the transitions. 

Evolved stars are generally strongest in the 1612-MHz transition and have 
a spectral shape dominated by two peaks with a separation of $30\pm10$
\kms\ (e.g. Fig. \ref{oh_spectra} left panel). The shape is due to the origin 
of the masers in an expanding circumstellar shell where the strongest 
amplification occurs radially. The two strong peaks then come from
the front and the back sides of the shell, and the radial velocity of
the star and the expansion velocity of the shell can be determined
directly from the central velocity of the double-peaked profile and the 
velocity interval covered by the emission. The
evolved-star masers vary in phase with the stellar luminosity 
variations. Superposed on this are peculiar variations due to flares or 
intervals of generally low emission levels.

All 4 ground-state lines can be observed towards SFRs, though the 
main-lines (1665 and 1667~MHz) are the most commonly observed.
Typically, SFR maser spectra are composed of narrow spectral components
($\leq$1-2~km~s$^{-1}$) which can spread over a wide range of velocity depending 
on the complexity of the SFR. The intensity of the individual components can 
vary by up to four orders of magnitude within the same SFR 
complex (e.g. Fig.~\ref{oh_spectra} right panel showing the rich 1665-MHz 
spectrum towards the Orion-KL SFR complex). 
Often, these spectra show substantial polarisation (including Zeeman 
pattern signatures) as well as substantial temporal variability 
(including flares) and periodic behaviours have also even been recorded 
towards class~II 6.7-GHz methanol masers 
(e.g. Goedhart, Gaylard \& van~der~Walt 2004). 

Masers originating from SFRs and supernova remnants (SNRs) are commonly 
referred to as ``interstellar masers''. This term shall be used hereafter, 
bearing in mind that we are focusing here on the evolved-star and SFR populations. The 
differentiation between SFRs and SNRs will be made to remove any ambiguity when 
appropriate.

\section{SKA survey framework}
\label{SKA Phase 1}
\subsection{Luminosity distributions of galactic OH masers}
\label{Galactic OH masers}
Currently >2000 stellar (Engels \& Bunzel, 2014; hereafter EB14) and
several hundred interstellar OH masers are known in the Milky
Way. Most were discovered by surveys with single-dish radiotelescopes
with typical survey limits of several hundred mJy. The most
comprehensive survey for stellar OH masers so far, is the ATCA/VLA OH
1612-MHz survey (Sevenster et al. 2001), covering the Galactic plane at 
$\mid l \mid \le 45^\circ$ and $\mid b \mid \le 3^\circ$. 
From this survey, we know that the number of detections is still increasing 
with increasing sensitivity. 

A series of surveys for interstellar masers
were made by Caswell, Haynes \& Goss (1980); Caswell \& Haynes 
(1983a\&b and 1987), Caswell (1998) and Caswell, Green \& Phillips 
(2013, 2014). They cover the Galactic plane to within 1 degree, in the range 
$233^\circ \le l \le 60^\circ$,  
and extend to masers slightly weaker than 1~Jy for the 1980s series of survey 
and $\sim$0.2~Jy for the more recent series. 
The ongoing Southern Parkes Large-Area Survey (SPLASH) for OH masers, 
which has a mean 5-sigma flux limit of 0.3~Jy, has already doubled 
the number of known masers for the area in common with previous 
surveys (Dawson et al. 2014).

\begin{figure}[h] 
\begin{minipage}[t]{15cm}
  \begin{minipage}[t]{7.5cm}
   \begin{flushleft}
      {\includegraphics*[width=5cm,angle=-90]{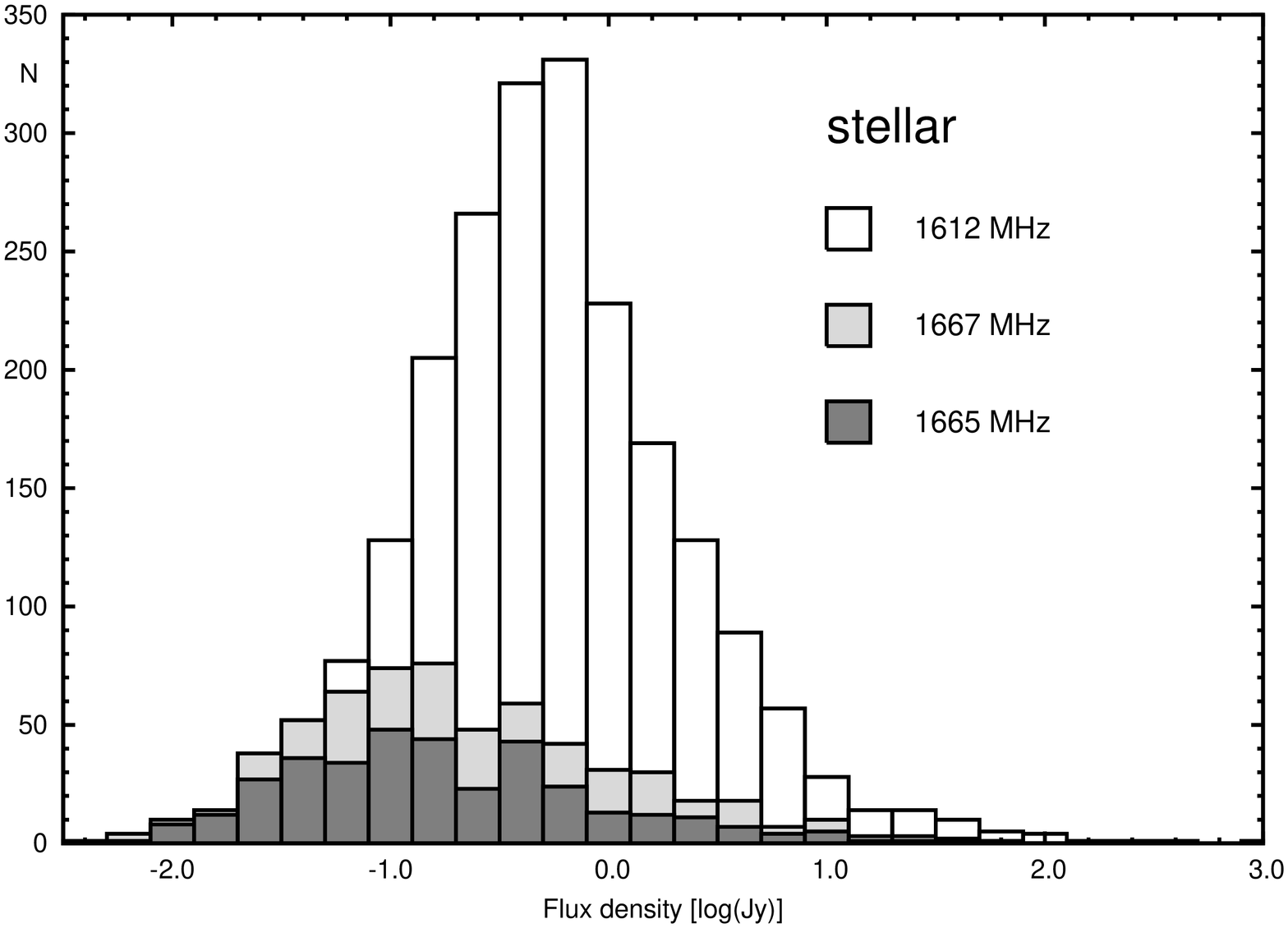}}
   \end{flushleft}
  \end{minipage}
  \begin{minipage}[t]{7.5cm}
   \begin{flushright}
      {\includegraphics*[width=5.25cm,angle=-90]{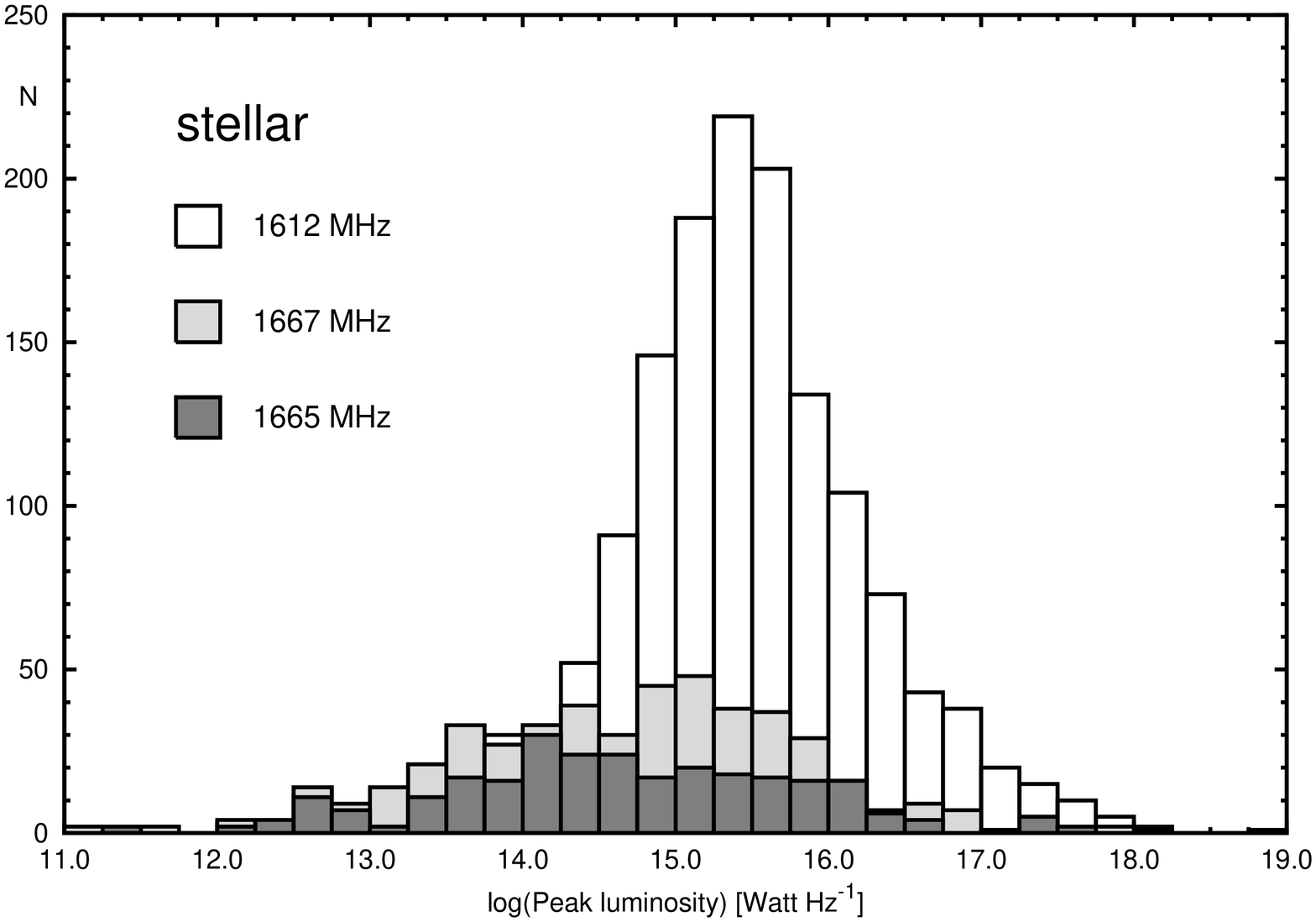}}
   \end{flushright}
  \end{minipage}
\hfill
  \begin{minipage}[t]{7.5cm}
   \begin{flushleft}
      {\includegraphics*[width=5cm,angle=-90]{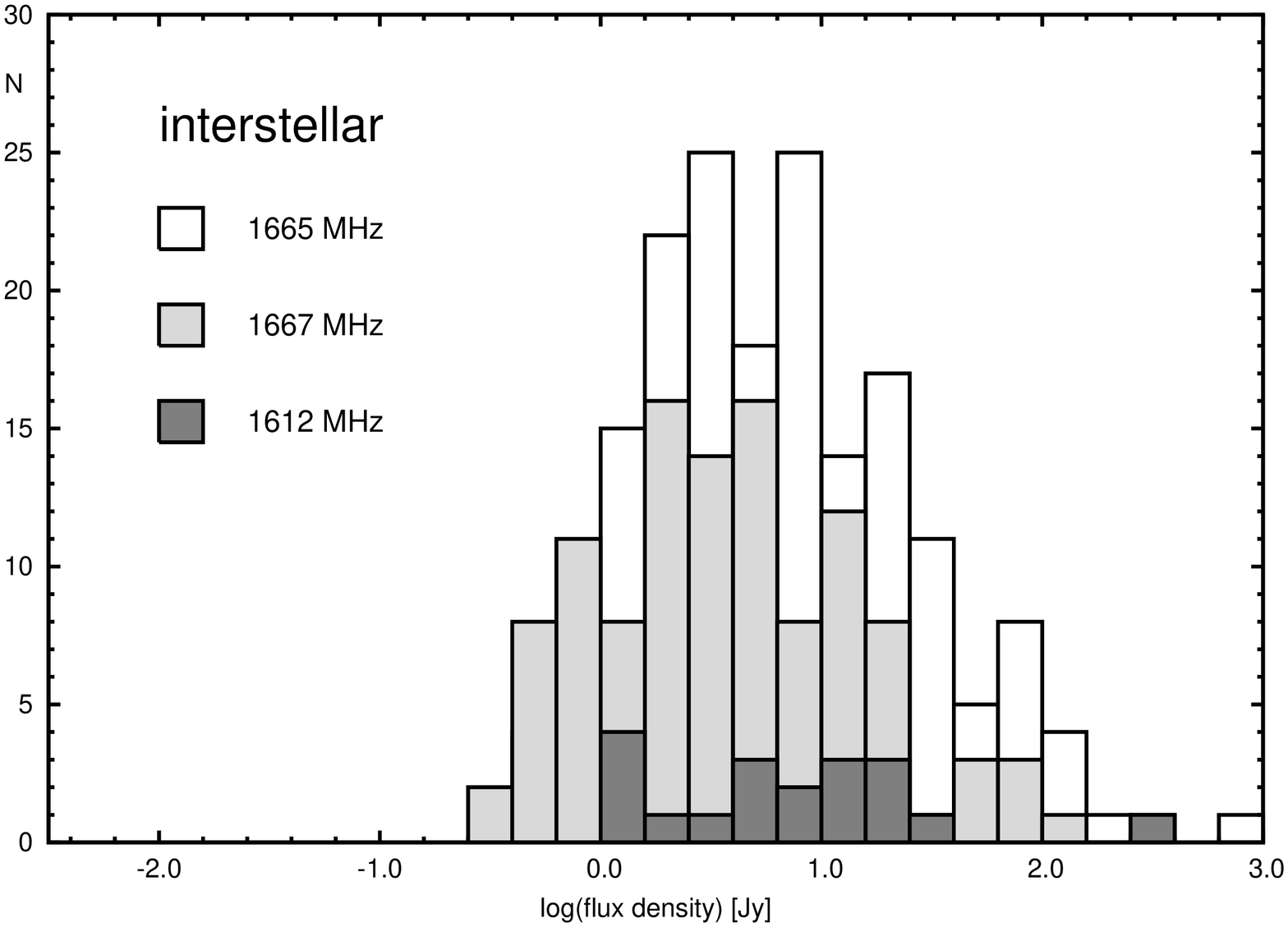}}
   \end{flushleft}
  \end{minipage}
  \begin{minipage}[t]{7.5cm}
   \begin{flushright}
      {\includegraphics*[width=5.25cm,angle=-90]{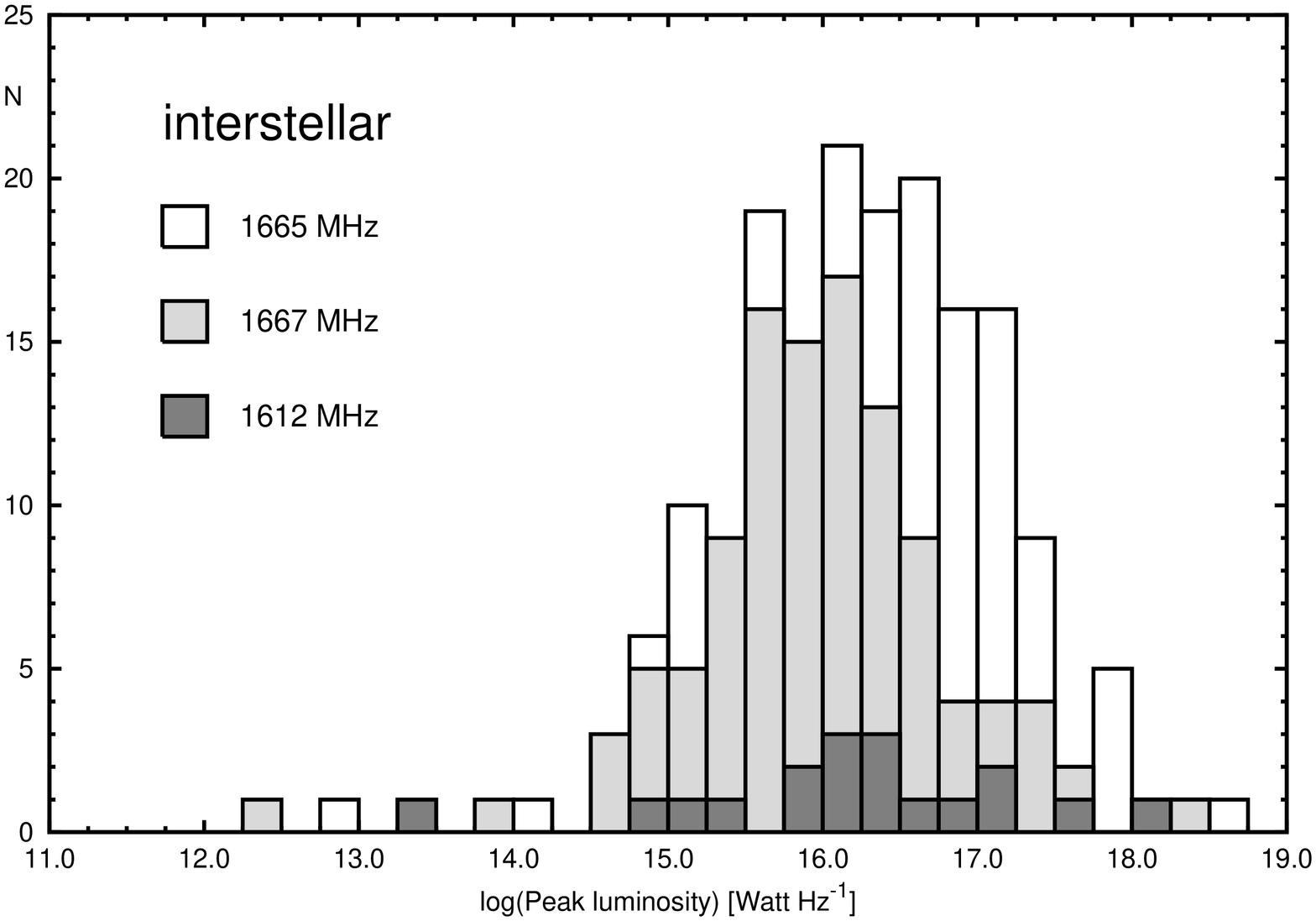}}
   \end{flushright}
  \end{minipage}
\end{minipage}
\caption{\small {\bf Left panel:} Distribution of peak flux densities
  for stellar masers (top) and for interstellar masers
  (bottom). Increasing shading denotes the 1612, 1667, 1665~MHz transition
  for stellar and 1665, 1667, 1612~MHz transition for interstellar masers.
  {\bf Right panel: } Distribution of corresponding
  peak luminosities assuming kinematic distances for stellar (top) and
  interstellar masers (bottom).}
\label{lable_ohdistri}
\end{figure}

To estimate the yields of more sensitive surveys, we have constructed
distributions of specific OH luminosities $L_\nu = f_{peak } \cdot 4
\pi D^2$, with $f_{peak }$ the peak flux density of the strongest
feature in the maser spectrum and $D$ the kinematic distance. The
distribution of flux densities and specific luminosities for stellar
sources, taken from EB14, are shown in the first row of
Fig.~\ref{lable_ohdistri}.  For the interstellar masers we used the
flux densities from the survey of Caswell and collaborators, 
and calculated specific luminosities using the kinematic distances 
provided by them. In case of ambiguity we used the distance quoted as 
'more likely'. The distributions obtained for the interstellar sources 
are shown in the second row of Fig.~\ref{lable_ohdistri}.

The flux distributions are peaking at $\sim0.5$~Jy (1612~MHz) for
stellar and at $\sim4$~Jy (1665/67~MHz) for interstellar masers,
reflecting the different sensitivities of the contributing
surveys. The luminosity distributions peak at 
$2 \times 10^{15}$~Watt~Hz$^{-1}$ and $\sim 10^{16}$ Watt\,Hz$^{-1}$ 
for the two different maser origins. 
The decrease in number at lower luminosities is very
likely due to incompleteness, and this will be probed by future
surveys.

\subsection{Implications of maser variability on surveys}
\label{Variability}
Due to the variability of the masers, single epoch surveys detect only
a fraction of the full population. Detection rates less than 50\% are common
for targeted surveys toward infrared selected samples of red giants.
The missing detections cannot be explained by sensitivity limitations
alone, but most likely are due to variability on a range of
timescales.  On timescales of years, the stellar masers vary in
response to the luminosity variations of the parent star. Repeated
observations usually produce detections for such cases. Part of
the O-rich population however are persistently not detected, although
'maser' and 'non-maser' AGB stars show no differences in their optical
or infrared properties. The fraction $\eta$ of 'maser' AGB stars
is dependent on infrared colour (e.g.  mass-loss rates) and ranges
between 10 and 60\% for a flux density limit of $\approx$50~mJy (Lewis, 1992).  

Either these 'non-maser' stars possess only low-luminosity masers or
the distribution of maser sites in their circumstellar shells is not spherically
symmetric, so that only those stars are detected in which the OH maser
emission is beamed toward Earth. Such issues are also valid for
interstellar masers.  For stellar masers, the beaming directions may 
change on timescales related to the crossing time through the shell
($\sim1000$~years). Therefore, we predict that the coming OH maser
surveys, which will be separated $\sim30$~years from the historical
ones, are also expected to discover new bright ($\gg1$~Jy) masers. The
rate of new discoveries in stars not detected previously are a direct
test of the timescales on which reconfigurations in the expanding
circumstellar shells may occur. 

\subsection{The Galactic OH maser population} \label{GalOH}
A rough estimate of the yield of future surveys for stellar OH masers
can be made for 1612~MHz using the model of Jackson et al. (2002) for
the Galactic distribution of AGB stars. This model is based on the
analysis of the Galactic distribution of $\approx10000$ IRAS sources
with colours appropriate for mass-losing AGB stars. It predicts
$\approx200\,000$ AGB stars in the Milky Way. 1612-MHz OH flux
densities can be estimated on the basis of dust mass-loss rates
$\dot{M}_{dust}$ (in \Myr) from  the relation 
$f_{OH} = 3.15\times10^{9} \, \dot{M}_{dust} \, (v_{exp} \, D^2)^{-1}$
(Zijlstra et al. 1996, taking into account the factor 2 
needed for a better fit of the LMC data as predicated by Marshall et al. 2004),
where $f_{OH}$ is the peak flux density in Jy, $v_{exp}$
the expansion velocity in km/s, and $D$ the distance in kpc. For a
given volume in the Galaxy, the space density of AGB stars was obtained
from the Jackson et al. model, and it was assumed that the stars have
a fixed IRAS colour distribution. The IRAS colours were converted to
$\dot{M}_{dust}$ using the pure silicate models of Groenewegen (2006),
which predict infrared colours for a range of gas mass-loss rates for
fixed bolometric luminosity $L = 3000$ \Lsun, $v_{exp} = 10$~\kms, and
a gas-to-dust ratio $\Psi = 200$. From the dust mass-loss rates, OH
flux densities were obtained as described above.

\begin{figure}[h] 
\begin{minipage}[t]{15cm}
  \begin{minipage}[t]{7.5cm}
   \begin{flushleft}
      {\includegraphics*[width=5cm,angle=-90]{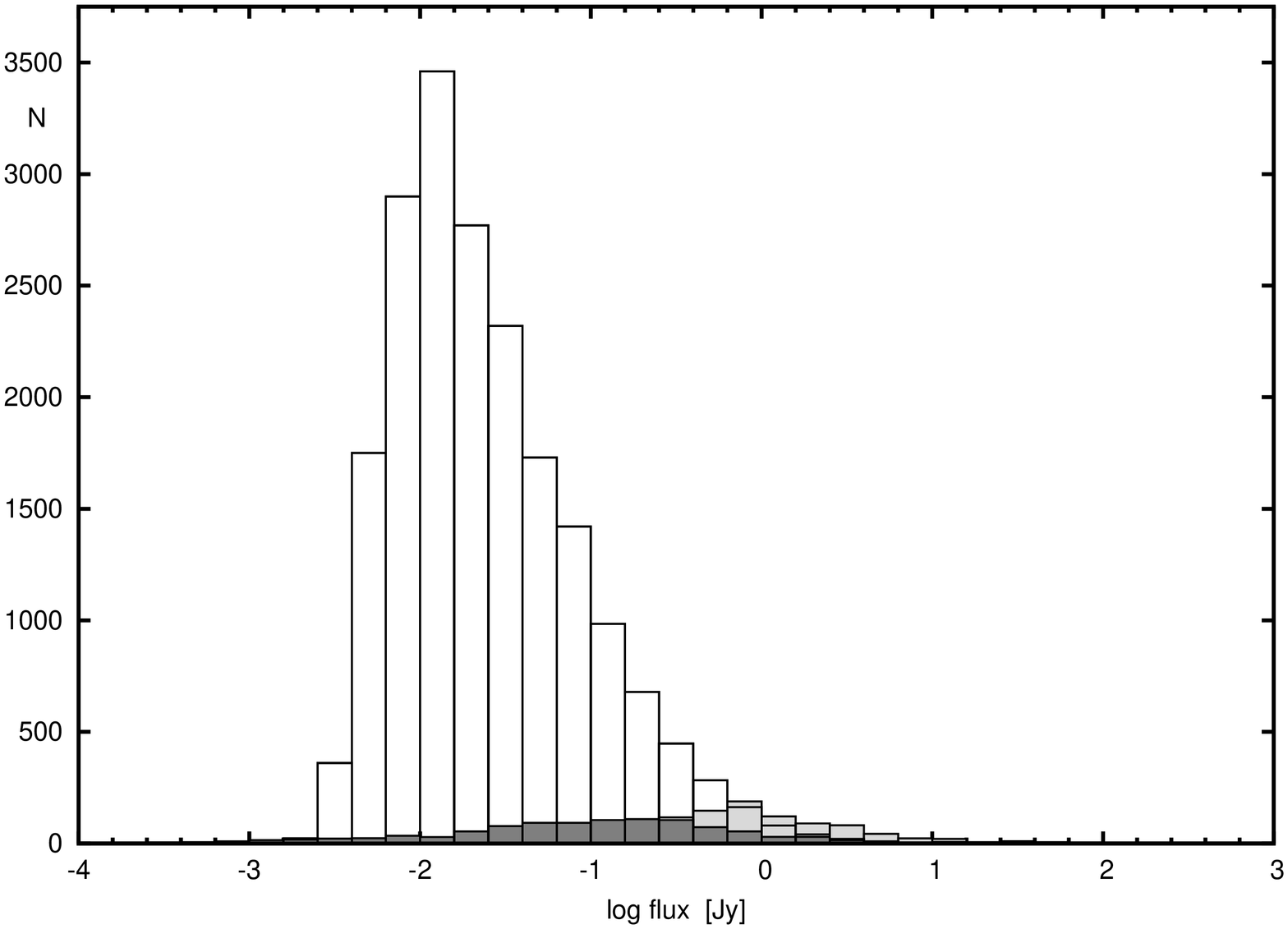}}
   \end{flushleft}
  \end{minipage}
  \begin{minipage}[t]{7.5cm}
   \begin{flushright}
      {\includegraphics*[width=5cm,angle=-90]{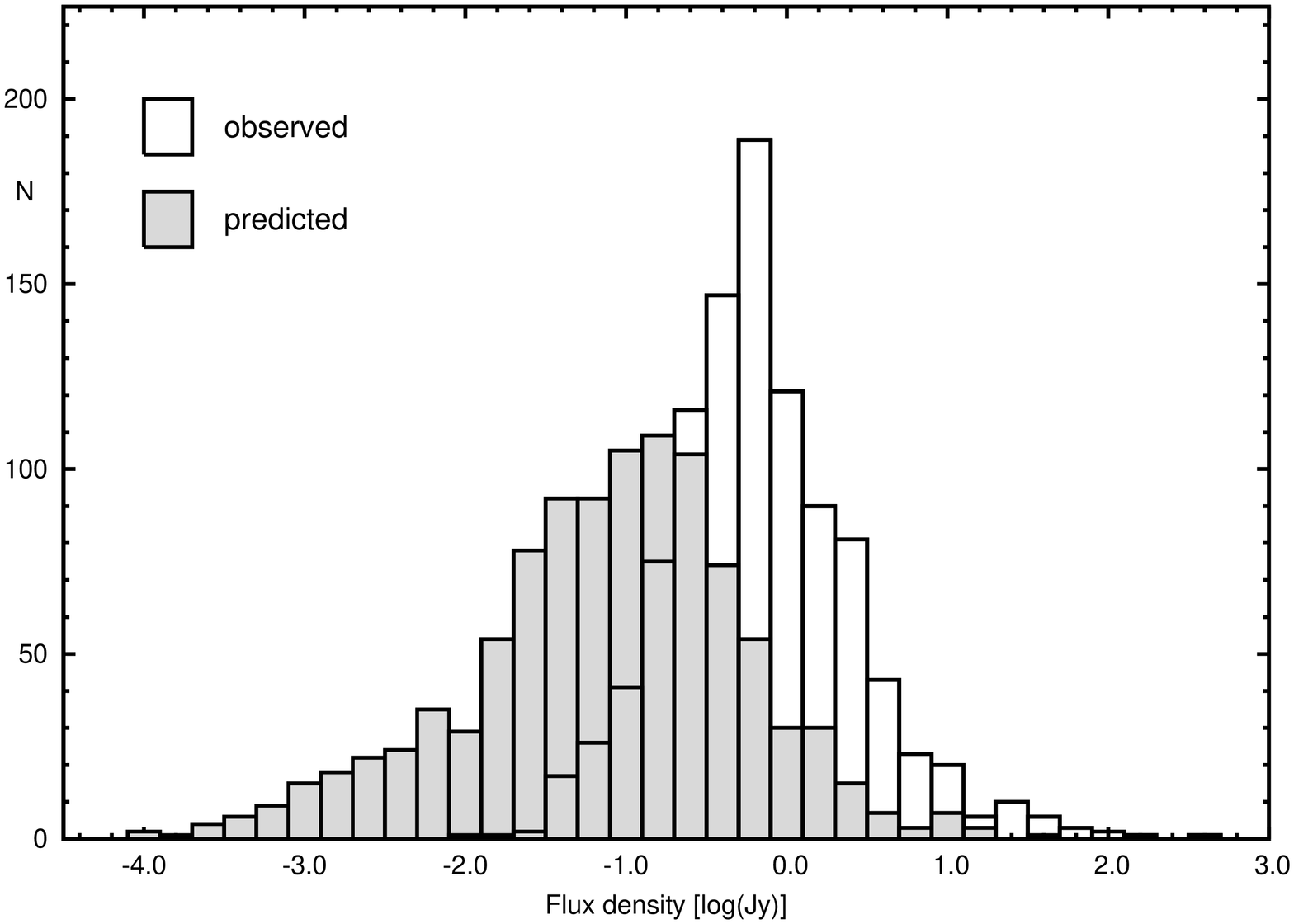}}
   \end{flushright}
  \end{minipage}
\end{minipage}
\caption{\small {\bf Left panel:} Predicted distribution of 1612-MHz
  stellar OH maser flux densities based on the model of the Galactic
  distribution of AGB stars of Jackson et al. (2002) and a relation
  between dust mass-loss rates and OH maser luminosities from Marshall et
  al. (2004). The shaded distributions are from the right panel.
  {\bf Right panel: } Observed stellar OH maser population in the 
                    front part of the solar Galactic hemisphere (that is up to 
                    a distance of $D \cdot cos(l) \le$~8.0~kpc) and its 
                    predicted counterpart in the anti-solar Galactic hemisphere
                    (see Sect.~3.2).
                    The observed population is taken from the EB14 catalogue.
}
\label{flux_histo_jackson_model}
\end{figure}

The actual numbers of OH masers predicted depend critically on the
distribution of the mass-loss rates (or IR colours) among the AGB star
population and also on the fraction $\eta$ of AGB stars actually
showing a maser with the predicted luminosity. The observed IRAS colour
distributions are biased towards higher mass-loss rates, because these
stars emit most of their energy in the mid-infrared. Corrections have
to be estimated therefore for low $\dot{M}_{dust}$. The range of
possible $\dot{M}_{dust}$ distributions and the fraction $\eta$ of OH emitting
AGB stars was constrained by the requirement that the model reproduces
the results of the survey of Sevenster et al. (2001), e.g. the number
of detections down to the survey limit of $\approx 150$~mJy and the
flux distribution.

The result for $\eta = 0.1$ is shown in
Fig. \ref{flux_histo_jackson_model} (left panel), giving
$\approx11\,000$ stellar OH 1612-MHz maser with $F_{OH} \ge 10$~mJy in
the Galactic plane ($\mid b \mid <3$ deg). 
  This value of $\eta$
  is at the lower end of the range observed by Lewis (1992), but
  includes also by default 'non-maser' carbon-rich AGB stars in
  the population. Higher estimates can be created by adopting a
  steeper colour distribution, demanding $\eta > 0.1$ to reproduce the
  observed high-luminosity masers. Accordingly, the prediction of the
  number of fainter masers will be scaled up.

In addition to the predicted number of stellar masers, an unknown
number of interstellar masers will be present in the Galactic plane.
Some of these masers will be detected prior to the SKA by surveys 
such as GASKAP. This survey aims to reach an rms for point sources of
1~mJy in the Galactic plane and of 0.5~mJy in the Magellanic Clouds.
However, the angular resolution will only be $\sim$20~arcsec  
(Dickey et al. 2013), and thus the masers at the lowest flux limits may 
not be distinguishable from thermal extended emission.
From a simple extrapolation of the number of identified maser sources
in the pilot SPLASH region ($\sim$200 sources within an area
10$^\circ$ width in Galactic Longitude), it is expected that at the 
SPLASH 5-sigma flux limit of 0.3~Jy several thousand maser sources 
are detectable. The number of interstellar masers detectable
down to  $F_{OH} \ge 10$~mJy is therefore at least as large as that  
predicted for the stellar masers, yielding a total number of detectable 
masers in the Galactic plane of $\ge 20000$.

\section{The impact of OH maser surveys with SKA1}
\label{Early Science survey}
Adopting the specific luminosity distributions from
Fig.~\ref{lable_ohdistri} as representative for the stellar and
interstellar maser luminosity distributions in the Galaxy, sensitivity
limits for an SKA survey to detect a major part of these maser
populations at different distances can be estimated. Adopting $3
\times 10^{14}$~Watt~Hz$^{-1}$ (at 1612~MHz for stellar and 1665/1667~MHz for
interstellar masers) as the lower limit for the maser
population to study, and a detection threshold of S/N=10, sensitivity
limits $F_{lim} =$~4, 1, and 0.1~mJy are required to probe the
population at distances of 8, 16 and 50~kpc respectively. These
correspond to the Galactic center, the anti-solar, and the 
Large Magellanic Cloud distances. \\

Most of the ground-state OH maser surveys undertaken in the Milky Way have been 
made with an angular resolution of typically 20~arcsec and GASKAP will deliver 
a similar resolution. 
A new SKA1 survey, with a resolving power 20 times that of GASKAP, 
will allow a direct comparison of the other side of the Galaxy, LMC \& SMC 
stellar and interstellar maser populations at resolution roughly similar to 
Galactic surveys. \\

The 3 ground-states lines at 1612, 1665 and 1667~MHz are simultanously 
covered by Band~2 with SKA1-MID and SKA1-SUR and the 4$^{th}$ OH ground-state 
line at 1720~MHz can be accessed with Band~3. While SKA1-SUR will offer 
an unprecedented survey speed, SKA1-MID, with a SEFD 4~times better than 
SKA1-SUR as well as an increase by a factor of 4 in the power of resolution
(0.22 compared to 0.9~arcsec respectively), will be better suited for 
investigating the structure of the maser emission to the same resolution as 
e.g. eMERLIN. SKA1-MID Band~5 would offer the ability of accessing the 6-GHz 
excited-OH maser transitions, as well as the class-II 6.7-GHz methanol maser 
transition associated exclusively with the formation process of high-mass stars.

In the following sections, OH maser surveys rendered possible with SKA1-MID and 
SKA1-SUR capabilities are discussed. They are summarised in 
Table~\ref{Table: SKA maser surveys}.

\begin{table}[hbt]
{\footnotesize
\caption{\small SKA1 maser surveys}
\label{Table: SKA maser surveys}
  \begin{tabular}{lllllll}
\hline
SKA1 Survey    & Sensitivity & Int. per$^1$ & Coverage                      & Pointing  & Total$^{1,2}$ & Aim \\
Name           & (mJy)       & pointing     &                               &     Nb    & time      &     \\
               &             & (hours)      &                               &           & (hours)   &     \\
\hline
SUR--All-Sky   &  4          & 0.11         & 21000~deg$^2$                 & 1300       & 143       & All-Sky shallow survey \\
SUR--Shallow   &  1.0        & 1.75         & $-60^\circ \le l \le +60^\circ$ &  30        & 52.5      & Anti-solar Galactic    \\
               &             &              & $ b \le 4^\circ$               &            &           & hemisphere maser pop.  \\
SUR--Deep      &  0.3        & 19           & $-60^\circ \le l \le +60^\circ$ &  30        & 570       & Seach for low    \\
               &             &              & $ b \le 4^\circ$               &            &           & luminosity pop. \\
SUR--LMC       &  0.1        & 175          &   11$\times$10~deg$^{2,3}$     &  9         & 1575      & LMC survey (3-sigma) \\
SUR--SMC       &  0.1        & 175          &   5.5$\times$3.5~deg$^{2,3}$   &  4         & 700       & SMC survey (3-sigma) \\
MID--MC        &  0.05       & 40           &                               &  100       & 4000     & LMC+SMC confirmation \\
               &             &              &                               &            &          & targeted survey (6-sigma) \\
MID--GAL       &  0.05       & 40           &   1~deg$^2$                   &  4         & 160       & Triangulum Survey \\
\hline
 \end{tabular} \\
{\bf 1}:         for a channel width of 5~kHz~$\Leftrightarrow$~$\sim$0.9~km/s;
                 SEFD$_{\rm MID}$=1.7 (cf. Table~1 of SKA--TEL--SKO--DD--001 (Dewdney et al. 2013); 
                     corresponding to ``combined'': 190 antennas + Meerkat); SEFD$_{\rm SUR}$=7.1 \\
{\bf 2}: FOV$_{\rm MID}$=0.5~deg$^2$; FOV$_{\rm SUR}$=18~deg$^2$ \\
{\bf 3}: from the NED (http://ned.ipac.caltech.edu/) \\
}
\end{table}

\subsection{OH masers in the anti-solar Galactic hemisphere}
\label{anti-solar hemisphere}
To estimate the likely range of flux densities of stellar OH masers in
the Milky Way located beyond the Galactic center (hereafter the
anti-solar Galactic hemisphere), we used all stellar masers with kinematic
distance estimates from EB14 with $D \cdot cos(l) \le$~8.0~kpc, e.g.,
the population of stellar masers in the solar Galactic hemisphere. 
We assumed that the population in the anti-solar Galactic 
hemisphere is a mirror of the front part population and calculated
their flux densities as if observed from the Sun.  The result is shown
in Fig.~\ref{flux_histo_jackson_model} (right panel), where the open bins 
represent the observed population and the shaded bins represent the predicted
population in the anti-solar Galactic hemisphere. The bulk of the predicted
population has flux densities between 10 and 500~mJy in good agreement
with the maximum of the flux density distribution obtained from
the prediction presented in Sect. \ref{GalOH}. This population can be detected 
in $\sim$50 hours with SKA1-SUR 
(cf. SUR-Shallow in Table~\ref{Table: SKA maser surveys}).

\subsection{Low-luminosity masers}
\label{Low-luminosity}
Surveys of increasing sensitivity in the Galaxy will not only extend
the distances out to which masers will be detected, but will also
extend the maser luminosity range towards lower levels than before. It
is unknown whether the specific luminosities of masers have lower
limits. A plausible assumption is that there is no such limit. 
A good illustration of this issue is the OH flaring Mira population 
(Etoka \& Le~Squeren 1997).  
$o$~Ceti is an ideal example amongst these stars, as it shows 
absence of detectable OH maser emission for decades followed by 
long-lasting OH flaring events (Etoka et al. 2010a). 
The mean intensity of the OH flaring emission towards $o$~Ceti is $\sim$2~Jy, 
corresponding to a luminosity of  $\sim 2 \times 10^{12}$~Watt~Hz$^{-1}$, 
which can be taken as a typical threshold for this type of stellar maser 
population.  The dichotomy 'maser' and 'non-maser' O-rich AGB stars
would then vanish with increasingly sensitive surveys.  In addition,
it seems also plausible that OH maser emission will be discovered in
classes of objects, which do not, or rarely possess high-luminosity
masers. One example are Planetary Nebulae (PNe). There are only six 
sources in which the presence of (1612~MHz) OH maser emission has been
confirmed (Uscanga et al. 2012). Another example is the so far unique 
detection of 1720-MHz maser emission from an evolved star (made
towards the post-AGB star OH\,9.097-0.392 by Sevenster \& Chapman 2001).
In post-AGB stars and PNe, the maser emission is
probably related to shock-excitation in the interface between a fast
post-AGB wind with a remnant slow AGB wind (Etoka et al. 2009). 
These excitation events might be short-lived, might lead to beamed 
emission or produce only low-luminosity maser emission.  It seems likely, 
that new sensitive surveys will discover many new masers in stars already 
evolved away from the AGB, allowing us to study the bipolar outflows emerging 
in this phase in more detail.  Towards SFRs, faint ground-state 1665/1667~MHz
OH maser from young stars which will not ionize their environment has
been found (e.g. the Turner-Welch object detected in the W3(OH) SFR complex by
Argon, Reid \& Menten 2003).
A $\sim$140-hours survey over the complete sky accessible by SKA1-SUR 
(cf. SUR-All-Sky in Table~\ref{Table: SKA maser surveys}) with a flux limit 
of 4~mJy will probe the presence of a population of low-luminosity masers 
(e.g. L$_{\rm OH} < 3 \times 10^{14}$ W\,Hz$^{-1}$) among evolved stars and SFRs 
in the solar vicinity, which previous surveys have only investigated for 
higher luminosity masers. A deeper blind search for such low-luminosity 
masers in the inner Galactic disk could be achieved within $\sim$570~hours
(cf. SUR-Deep in Table~\ref{Table: SKA maser surveys}). \\

\subsection{Galactic analog OH masers in galaxies of the Local Group}
\label{Galactic analog in the Local Group}
Late-type star evolution and the star formation process may differ in
galaxies with other metallicity environments. The extension of the
Galactic maser research to galaxies in the Local Group will provide
another tool to study these differences. Due to their proximity, the
Magellanic Clouds with their low-metallicity environment are the first
galaxies to address.

Only a small number of maser sources with luminosities similar to 
their Galactic analogs are known outside the Milky Way. In the LMC, 
ten 1612-MHz OH masers with peak-flux densities 17-600~mJy are known in 
evolved stars (Marshall et al. 2004). 
These low-metallicity, high-mass LMC stars have contributed significantly 
to our understanding of the expansion velocities as indicators of metallicity.
A few searches for ground-state OH maser of interstellar origin have 
been made in the 1980's and 1990's. A handful of 1665-MHz maser sources 
(sometimes accompanied with fainter 1667~MHz) have been detected towards 
the LMC (Brooks \& Whiteoak 1997).  A similar number of detections of 1720-MHz
maser emission originating from SNRs have also been found towards the
LMC (Brogan et al. 2004). Regarding the SMC, to date, OH masers have not been 
detected. Furthermore, maser emission of evolved-star origin has not been 
detected so far, and the number of maser sources of SFR 
origin detected is very low (in total 6 H$_2$O masers of SFR origin: 
Scalise \& Braz 1982; Breen et al. 2013; also, the systematic survey done 
by Green et al. (2008) failed to detect any class~II 6.7-GHz methanol masers).

\begin{figure}[h] 
\begin{minipage}[t]{15cm}
  \begin{minipage}[t]{7.5cm}
   \begin{flushleft}
      {\includegraphics*[width=5cm,angle=-90]{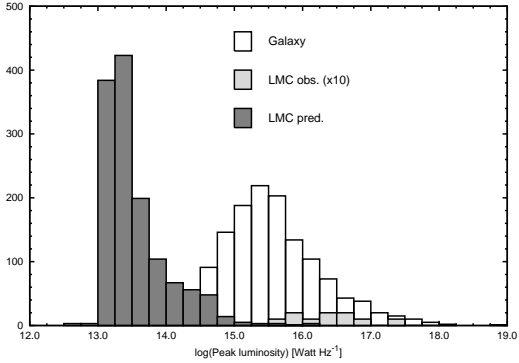}}
   \end{flushleft}
  \end{minipage}
  \begin{minipage}[t]{7.5cm}
   \begin{flushright}
    \caption{\small Predicted specific luminosity distribution for
                    1612-MHz OH masers in the LMC. 
                    The prediction is compared to the luminosities
                    of the observed masers in the LMC (numbers are
                    multiplied by 10) and the distribution of known
                    masers in the Galaxy from the EB14 catalogue.}
    \label{lumino histo 1612 lmc}
   \end{flushright}
  \end{minipage}
\end{minipage}
\end{figure}

With the sensitivities provided by SKA1, the options for maser
research in the Magellanic Clouds may dramatically increase.
To verify our ideas about Galactic structure evolution, it would be of
great importance to compare the populations of stellar masers in the 
Galaxy with those in other galactic environments.
Currently known stellar masers in the LMC were discovered by targeted searches
towards luminous and mid-infrared bright sources. Their specific
luminosities $L_{OH}$ cover the range $15.50 < \log L_{OH} < 17.50$,
which corresponds to the bright tail of the luminosity distribution in
the Galaxy (cf. Fig.~\ref{lumino histo 1612 lmc}). An inventory of
evolved stars in the LMC ($\sim50$ deg$^2$) and the SMC ($\sim30$
deg$^2$) has been provided by the SAGE-Spitzer survey (Meixner et al. 2006,
Gordon et al. 2011) in the infrared. In the LMC, depending on the
photometric classification according to chemistry, 20000--25000 of
them are expected to be oxygen-rich (Riebel et al. 2012). For most
stars, the dust mass-loss rates given by Riebel et al. are low
($\dot{M}_{dust}<5\times10^{-10}$ \Myr) and their circumstellar shells 
do not qualify as OH maser hosts. 
To estimate the OH maser luminosities of the $\sim1300$ 
stars with higher dust mass-loss rates we used the relation from 
Sect.~\ref{GalOH} relating $\dot{M}_{dust}$ with 1612-MHz OH maser flux 
densities $f_{OH}$. Assuming $v_{exp} = 14$ \kms\ and D~=~50~kpc, the 
luminosity distribution shown in Fig.~\ref{lumino histo 1612 lmc} is 
obtained. The result indicates that most bright masers in the LMC are 
already known, and the luminosities of the masers to be discovered will 
be lower than that of the bulk of the known Galactic masers. The reason 
for this is the classification of almost all stars with high mass-loss rates 
(``extreme AGB stars'') as C-stars by the SAGE-Spitzer survey. 
The validity of this photometric classification will be directly probed 
by an OH maser survey of the LMC covering the full AGB population.
 
As expected, the number of masers in Fig.~\ref{lumino histo 1612 lmc}
increases for lower luminosities, following the increase of the number
of AGB stars with decreasing mass-loss rates. With a survey limit of
0.1~mJy and S/N=3, $\approx200$ new stellar 1612-MHz OH masers with
$L_{OH} > 10^{14}$ Watt~Hz$^{-1}$ will be detected.  These numbers
should be seen as upper limits, if detection rates as in the Galaxy
apply. On the other hand, more 1612-MHz OH masers might be found if 
there are more O-rich AGB stars among the ``extreme AGB stars'' than 
expected. Furthermore, since (preferentially main-line) masers are
commonly present in the vicinity of Galactic YSOs, they are probably 
present around their LMC analogs. A mid-IR photometry search for 
high- and intermediate-mass YSOs in the LMC has 
been made by Gruendl \& Chu (2009) yielding $>1000$ sources.
The number of OH masers detectable in the LMC with a 0.1~mJy survey 
limit is therefore well above 200.

The number of O-rich evolved stars qualifying as potential OH maser
hosts in the SMC is $>$2000 (Boyer et al., 2011), about 10\% of the
LMC. Assuming, that the number of YSOs is also 10\% of the corresponding 
number in the LMC, and allowing for the larger distance (D = 60~kpc), 
about two dozen new OH masers might be found, adopting the same survey 
limit as in the LMC. They will be the first ever detected in this galaxy.
Both Magellanic Clouds can be surveyed in $\sim$2300~hours for the 
0.1~mJy flux limit required 
(SUR-LMC, SUR-SMC in Table~\ref{Table: SKA maser surveys}). 
Follow-up observations to improve the spatial resolution and the 
signal-to-noise ratio of the most interesting new discoveries are 
foreseen using SKA1-MID, adopting 100 pointings as a figure of merit 
(MID-MC in Table~\ref{Table: SKA maser surveys}). 

The recent interest in the evolved stellar population and in the
mass-loss and age-metallicity relation in the Sagittarius Dwarf
Spheroidal (Lagadec et al. 2008 \& 2009), located at only 25~kpc 
midway between the Galaxy and the LMC, has prompted searches for OH 
masers in this dwarf galaxy.  However, the current sensitivity limits 
are barely sufficient to perform statistical comparisons with evolved 
stars in the Galaxy and other nearby different metallicity systems. Such
studies clearly have to wait for SKA.

Triangulum (M33) is the third largest member of the local group, and
with $\delta = 30.6^\circ$ is accessible with the SKA. Three H$_2$O
masers are known in this galaxy (Brunthaler et al. 2006), while searches 
for methanol masers (Goldsmith et al. 2008) and OH masers
(Fix \& Mutel 1985; Baan, Haschick \& Henkel 1992) have to date been 
unsuccessful. The limiting sensitivity of the OH observations was 
$\approx15$~mJy ($5\sigma$), which sets an upper limit of 
$1.5\times10^{18}$~Watt~Hz$^{-1}$ on the luminosity of potential M33 OH 
masers, adopting a distance of $\approx880$~kpc.  The non-detection of OH 
masers in previous observations is not surprising, as only a small number of 
Galactic OH masers have luminosities in excess of this limit.
We expect however that the brightest interstellar and the bright stellar 
OH masers, with $L_{OH} > 5 \times 10^{15}$ Watt~Hz$^{-1}$, among the more than 
10\,000 AGB stars identified by Cioni et al. (2008) in M33 will be detectable 
with a survey (MID-GAL in Table~\ref{Table: SKA maser surveys}), having
the same sensitivity to that proposed for the LMC+SMC follow-up 
observations. \\

The study of OH masers with SKA in galaxies beyond the Local Group will focus on
OH megamaser emission. This emission is 10$^8$~times more luminous than Galactic 
analog OH masers, and is observed towards luminous and ultra-luminous infrared galaxies.
These megamaser-hosting galaxies are either mergers or show evidence of 
interaction with other galaxies and are associated with a burst of star 
formation. Hence, megamasers are useful probes of the conditions where star 
formation is taking place in such galaxies as well as allowing 
extragalactic magnetic field measurements
(McBride, Heiles \& Elitzur 2013, Robishaw, Quartaert \& Heiles 2008, 
cf. also Beswick et al. \& Robishaw et al. in these proceedings).

\section{Survey applications}
\label{Survey applications}

\subsection{Polarisation \& magnetic fields}
\label{Polarisation and Magnetic fields}
OH masers, characterised by their spectral narrowness and their high 
sensitivity to magnetic fields, can be strongly polarised and commonly exhibit 
Zeeman splitting. Hence, they are a particularly useful tool for polarisation 
properties studies and retrieving the magnetic field structure and strength of 
the medium they are probing. 
They can be used for this purpose towards both evolved-stars 
(Amiri, Vlemmings \& van Langevelde 2011; Etoka \& Diamond 2010b) and SFRs 
(Caswell, Hutawarakorn  \& Reynolds 2011a,b; Asanok et al. 2010; 
Wright, Gray \& Diamond 2004a,b; Hutawarakorn et al. 2002).
Not only can the Zeeman splitting be used to explore the magnetic fields from 
individual objects, but also as a whole throughout the Milky Way and the 
galaxies of the Local Group 
(Green et al. 2012, cf. also Robishaw et al. in these proceedings).

\subsection{Stellar kinematics}
\label{Stellar kinematics}
Similar to SiO masers, a large variety of stellar populations host OH masers: 
evolved stars in the Galactic thick disk, bulge, and globular clusters in the 
halo as well as young stellar clusters in the Galactic thin disk 
(Deguchi et al. 2004). A large sample of OH masers will reveal the 
velocity fields of these Galactic components.
The high luminosity of the masers together with the radial velocity
information provided, allow the study of the stellar kinematics 
in the various Galactic components (yielding for example evidence for a 
bar in the Milky Way: Habing et al. 2006).
The information on the kinematics, supplemented with the age and metallicity
of the stars, will point to common origins and clues for Galactic 
evolution models. These studies will be greatly enhanced by increasing 
numbers of southern sources.

\subsection{VLBI \& astrometry}
\label{VLBI and Astrometry}
Among evolved stars, OH/IR stars are sizeable objects of typically 
10000~AU when mapped in the ground-state OH (Etoka \& Diamond 2004, 2010b), 
and so are SFR complexes (Cohen et al. 2006). This typical size corresponds 
to 1.25~arcsec at 8~kpc and 0.2~arcsec at 50~kpc.  With an achievable
resolution of 0.22~arcsec, SKA-MID will be able to map a good fraction
of the OH/IR stars and SFR complexes detected at Galactic-center
distances and beyond, as well as the strongest and most extended
objects in the LMC. 
To derive distances, VLBI-Astrometry (Vlemmings \& van Langevelde 2007) 
or the ``phase-lag technique'' can be used (Engels et al. 2014).
Since OH/IR stars can be found in all the Galactic components 
(cf. the previous subsection~\ref{Stellar kinematics}), as opposed to 
SFRs, and provided that the OH circumstellar shell is extended, the distances 
of these various components (and hence a better presentation of the galactic 
structure) can potentially be inferred using this method. 

Admittedly, extended OH/IR stars can be significantly resolved out with 
baselines longer than 1000~km (Imai et al. 2013). Since the resolving 
power increases with frequency, higher frequency masers (i.e. in Band~5) in 
VLBI in-beam mode will allow high resolution astrometry and hence distance 
determinations with the possibility of a 3D mapping of the structure of our 
Galaxy but also that of the LMC and SMC (cf. Green et al. in these proceedings).

\subsection{SNR \& the Galactic center circumnuclear Disk}
\label{SNR GCN disk}

Whereas the 1612-MHz OH masers are predominantly seen in
circumstellar environments and the mainline 1665- and 1667-MHz OH
masers are signposts for SFRs, the collisionally pumped 1720-MHz
OH maser is typically seen as the only OH transition outlining
shocked environments. Small shocks, like in the formation of
massive stars and in post-AGB outflows are currently difficult to
detect and thus limit our understanding of these stellar
evolutionary phases. However, the more energetic shocks generated
by SNRs plowing into dense (molecular) clouds are readily
recognised by bright 1720-MHz maser emission observed at the
interaction regions (Claussen et al. 1997, Frail et al. 1996). 
With more sensitive observations available, less bright masers appear
and make it possible to probe shocks in different evolutionary stages
of stars. The construction of 3-dimensional velocity models of expanding 
SNRs and of dynamical structures like the circumnuclear disk in the 
Galactic center will be possible (Sjouwerman \& Pihlstrom 2008).

\subsection{Synergy between HI and OH surveys}
 \label{Synergy HI - OH}
Recent studies have revealed weak extended HI shells around evolved stars 
(Libert, G\'erard, \& Le Bertre 2007). While GASKAP will be able to provide a 
catalogue of such shells, the SKA higher angular resolution and sensitivity 
is needed for the study of these HI shells in greater detail. This will 
provide extra information on the mass-loss history and on the way stellar 
matter, enriched in elements produced in the stellar interior, 
is re-injected into the interstellar medium. 

Since SFRs are converging and dispersing points of matter in the 
interstellar space, combining the OH maser distribution and the map 
of HI emission will provide a panoramic view of the life-cycle of matter 
in the Galaxy. GASKAP will pioneer this aspect 
through simultaneous surveys of OH and HI emission. SKA1 will allow 
a better correlation between OH masers and HI clumps by resolving the 
smaller structures of HI directly associated with the OH maser 
sources.

\section{Prospects of research with the full SKA enhanced capabilities}
\label{SKA Phase 2}
Because of the complex non-linear nature of maser pumping, there is no simple 
means of inferring the physical conditions from observations from the maser 
intensity itself. Nonetheless, maser transitions are inverted for a range of 
physical conditions. Thus, combining various maser species and within the same 
species different transitions to study SFR regions is quite powerful as a 
probe of e.g. local density and temperature variations down to a few hundreds 
of AU (Etoka, Cohen \& Gray 2005). Hence the combined information can give us 
access to the different physical components structure (i.e., outflows or disks) 
around a given YSO (Etoka, Gray \& Fuller 2012).

As mentioned in section~\ref{Early Science survey}, Band~2, 3 \& ~5 will 
allow access to a wide range of maser transitions. 
Since different maser transitions have different sensivities, or 
dependencies on the physical conditions, the presence or absence of the 
various transitions is expected to change as the star formation region evolves. 
These relationships can only be determined through statistical population 
studies of the different maser transitions (e.g. Breen et al. 2010).  The 
ability of the SKA and its pathfinder instruments to make rapid, 
sensitive surveys of the Galactic Plane will open up new opportunities to 
undertake such studies on scales which have not previously been possible. 

OH masers are already known to be present towards a handful of PPNe-PNe in 
the Galaxy, but it could be that faint OH masers (e.g., from AGB OH-shell 
remnants) towards these objects have not been detected in previous surveys 
due to sensitivity. 
Similarly, excited OH maser has also been found towards a handful of evolved 
objects (Desmurs et al. 2010). 

The sensitivity and frequency range covered by the full SKA,  will allow 
us to increase the statistics on these ``rare events'' through a systematic 
survey, and identify when in the evolutionary stage this occurs and probe for 
other possible ``new classes of masers''. 

\acknowledgments
This Chapter is dedicated to the late Jim Caswell, whose review helped to 
improve the text considerably. His extensive and systematic research on masers 
in star-forming regions was invaluable to evaluate the prospects of SKA for 
maser research.

\def\newblock{\hskip .11em plus .33em minus .07em}

\end{document}